\documentclass[twocolumn]{aastex62}

\newcommand{\MS}{\ifmmode{\,}\else\thinspace\fi{\rm M}\ifmmode_{\odot}\else$_{\odot}$\fi}
\newcommand{\LS}{\ifmmode{\,}\else\thinspace\fi{\rm L}\ifmmode_{\odot}\else$_{\odot}$\fi}
\newcommand{\RS}{\ifmmode{\,}\else\thinspace\fi{\rm R}\ifmmode_{\odot}\else$_{\odot}$\fi}
\newcommand{\teff}{\ifmmode T_{\rm eff}\else$T_{\rm eff}$\fi}
\newcommand{\Ke}{\ifmmode{\,}\else\thinspace\fi{\rm K}}

\received{\today}
\revised{\today}
\accepted{\today}

\submitjournal{ApJ}

\shorttitle{First-Overtone Type II Cepheids}
\shortauthors{Soszy\'nski et al.}

\begin{document}

\title{Type II Cepheids Pulsating in the First Overtone from the OGLE Survey}

\author{I. Soszy\'nski}
\affiliation{Warsaw University Observatory, Al. Ujazdowskie 4, PL-00-478 Warszawa, Poland}
\email{soszynsk@astrouw.edu.pl}

\author{R. Smolec}
\affiliation{Nicolaus Copernicus Astronomical Center of the Polish Academy of Sciences, ul. Bartycka 18, PL-00-716 Warszawa, Poland}

\author{A. Udalski}
\affiliation{Warsaw University Observatory, Al. Ujazdowskie 4, PL-00-478 Warszawa, Poland}

\author{P. Pietrukowicz}
\affiliation{Warsaw University Observatory, Al. Ujazdowskie 4, PL-00-478 Warszawa, Poland}

\begin{abstract}

We report the discovery of the first type II Cepheids (BL~Herculis
stars) pulsating solely in the first overtone. We found two such
objects among tens of millions of stars regularly observed by the OGLE
survey in the Large Magellanic Cloud. Our classification and the
pulsation mode identification is based on the position of these stars
on the period--luminosity and color--magnitude diagrams and on the
light curve analysis. We discuss why single-mode first-overtone BL~Her
pulsators must be very rare. For the two discovered stars we present
non-linear models that successfully reproduce their light
variation. These models indicate that both first-overtone pulsators
should be more massive than it is typically assumed for BL~Her stars,
i.e. their masses should be above $0.75\MS$. However, the higher mass
requires higher luminosity to match the observed periods of the stars,
which is inconsistent with observations.

\end{abstract}


\section{Introduction} \label{sec:intro}

The standard model of stellar pulsations \citep{eddington1926}
predicted that stars may pulsate not only in the fundamental mode but
also in the overtone modes. \citet{edgar1933} was the first who
theoretically studied this idea in detail, while
\citet{schwarzschild1940} first suggested that RR Lyrae stars of type
c \citep{bailey1902} are the first-overtone pulsators. The physical
nature of the so-called sinusoidal or s-Cepheids has been a matter of
debate for years
\citep[e.g.][]{arp1960,ivanov1976,connolly1980,antonello1990,gieren1990}
until large-scale microlensing surveys: EROS \citep{beaulieu1995},
MACHO \citep{alcock1995} and OGLE \citep{udalski1999} presented large
samples of classical Cepheids in the Magellanic Clouds and
unambiguously proved that s-Cepheids are the first-overtone
pulsators. Modern catalogs of classical Cepheids contain single- and
multi-mode variables with the first, second and even the third
overtone excited \citep[e.g.][]{soszynski2015}.

In the virtually complete collection of Cepheids and RR Lyr stars
(collectively referred to as classical pulsators) in the Magellanic
Clouds published by the OGLE survey
\citep{soszynski2015,soszynski2017a}, the overtone pulsators
constitute a significant fraction of the total population. About 28\%
of all RR~Lyr stars, 31\% of anomalous Cepheids, and 46\% of classical
Cepheids have the overtone modes excited (including multi-mode
pulsators). Type II Cepheids are an exception in the family of
classical pulsators -- until recently all known variables of this
class were pure fundamental-mode pulsators.

\begin{figure*}
\plotone{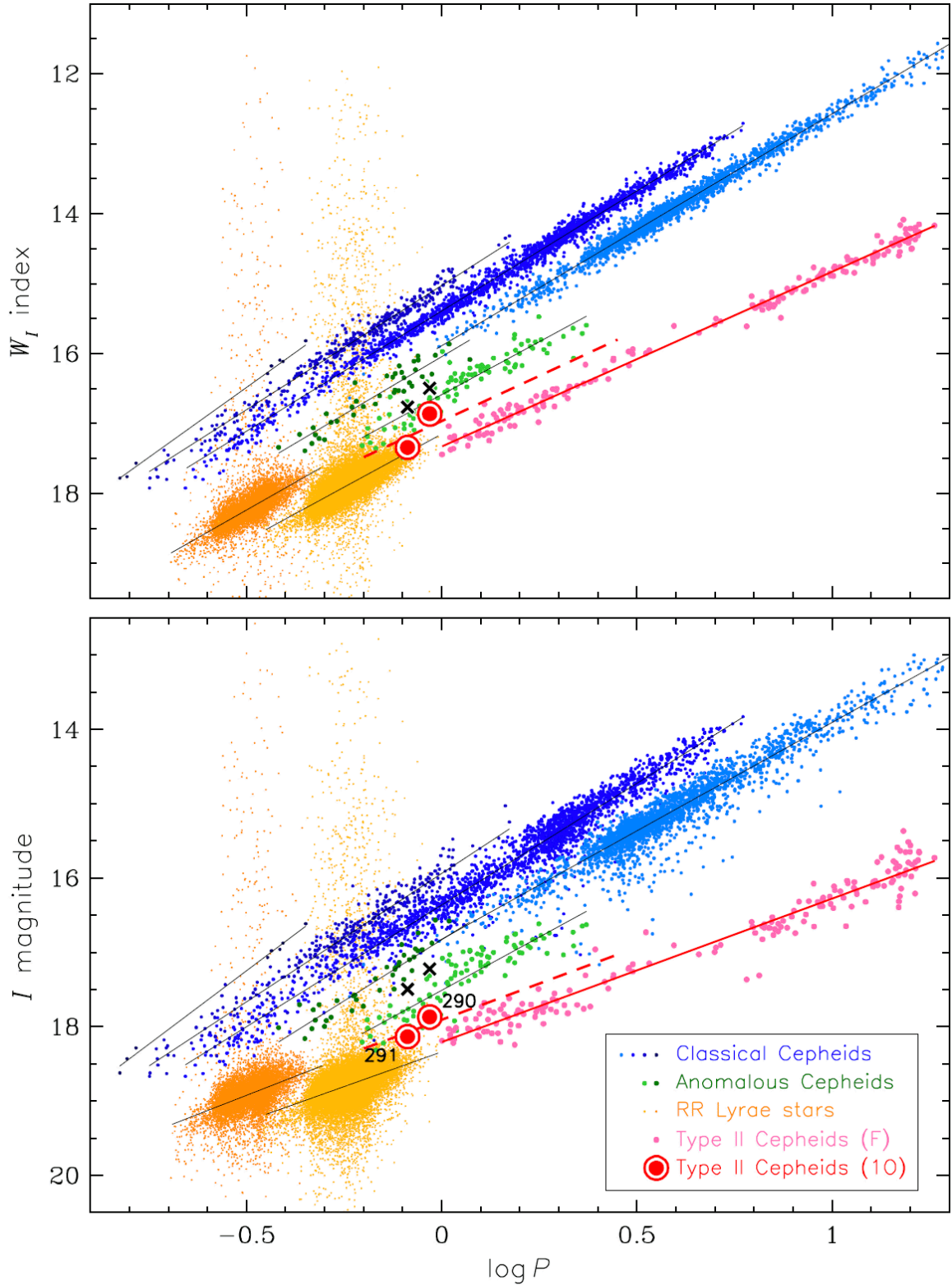}
\caption{Period--Wesenheit index (upper panel) and period--luminosity
  (lower panel) diagrams for classical pulsators in the LMC. Blue
  points represent classical Cepheids, green points -- anomalous
  Cepheids, pink points -- fundamental-mode type II Cepheids, and
  orange points -- RR Lyr stars. Lighter and darker colors indicate
  fundamental-mode and overtone-mode pulsators, respectively. Solid
  lines show least-square fits to the relations (red lines for type II
  Cepheids and black lines for other classical pulsators). Dashed red
  lines show expected relations for the first-overtone BL~Her
  stars. Red circles mark our two candidates for the first-overtone
  BL~Her stars. Black crosses correspond to the best matching
  pulsation models for OGLE-LMC-T2CEP-290 and OGLE-LMC-T2CEP-291 (see
  Section~\ref{sec:models}).\label{fig1}}
\end{figure*}

The first two double-mode type II Cepheids pulsating simultaneously in
the fundamental and first-overtone modes were discovered by
\citet{smolec2018}, who identified them among nearly one thousand type
II Cepheids detected by the OGLE survey in the Galactic bulge
\citep{soszynski2017b}. These two unique objects have
(fundamental-mode) periods equal to 1.04~d and 1.18~d, so they belong
to the shortest-period subclass of type II Cepheids -- BL~Herculis
stars -- just above the conventional borderline (1~d) between RR Lyr
stars and type II Cepheids. The ratios of the first-overtone to the
fundamental-mode periods for these two stars are almost identical and
are equal to about 0.705. Recently, \citet{udalski2018} reported the
discovery of two other double-mode BL~Her stars in the Galactic bulge
and disk.

Inspired by these discoveries, we decided to search the OGLE
photometric databases for single-mode first-overtone BL~Her stars. The
best environment to perform such a search is the Magellanic System, in
particular the Large Magellanic Cloud (LMC), because it contains rich
populations of spatially resolvable stars at approximately the same
distance from us. This allow us to precisely locate pulsating stars on
the period--luminosity (PL) diagram to find out in which mode they
pulsate. The long-term OGLE photometry is useful for studying the
light curve morphology which is also an invaluable discriminator
between fundamental and overtone pulsators.

\section{Observational Data and Candidate Selection} \label{sec:obs}

The OGLE database contains light curves of over 75~million stars in
the Magellanic Clouds. These data have been collected with the 1.3-m
Warsaw Telescope located at Las Campanas Observatory in Chile. The
observatory is operated by the Carnegie Institution for Science. Since
2010, the Warsaw Telescope has been equipped with the 256 Megapixel
mosaic CCD camera with a field of view of 1.4~square degrees at a
scale of 0.26\arcsec/pixel. OGLE uses the {\it V}- and {\it I}-band
filters from the Johnson-Cousins photometric system, but most of the
observations (typically 700 points per star) have been carried out
with the {\it I} passband. Details of the OGLE instrumental setup,
data reduction and calibration can be found in \citet{udalski2015}.

Candidates for the first-overtone BL~Her stars have been sought among
sources observed since 2010 by the OGLE-IV survey. First, we used the
{\sc Fnpeaks}
code\footnote{http://helas.astro.uni.wroc.pl/deliverables.php?active=fnpeaks}
to compute the Fourier amplitude spectra for all {\it I}-band light
curves obtained by OGLE in the Magellanic System to determine periods
with corresponding amplitudes and signal-to-noise ratios.

Second, we defined PL relations obeyed by hypothetical first-overtone
type II Cepheids, separately for the LMC and Small Magellanic Cloud
(SMC). We used the $\log{P}$ vs. $W_I$ and $\log{P}$ vs. {\it I}-band
luminosity planes, where $W_I$ is the reddening-independent Wesenheit
index, defined as $W_I=I-1.55(V-I)$, and $I$ and $V$ are the
intensity-averaged mean magnitudes in these filters. In
Figure~\ref{fig1}, we present the period-Wesenheit (PW) and PL
diagrams for classical pulsators in the LMC
\citep{soszynski2017a,soszynski2018}: classical Cepheids (blue
points), anomalous Cepheids (green points), type II Cepheids (pink
points), and RR Lyr stars (orange points). Lighter and darker colors
indicate fundamental-mode and overtone-mode pulsators,
respectively. The dashed red lines show the expected PW and PL
relations for the first-overtone BL~Her stars. These lines have the
same slopes as the relations for the fundamental-mode type II Cepheids
(plotted with the solid red lines) and are shifted toward the shorter
periods, corresponding to the period ratio of 0.7 between both modes.

Third, we visually inspected all light curves in both Magellanic
Clouds meeting the following criteria: (i) the primary period in the
range 0.7--3~d (corresponding to 0.7 of the period range of
fundamental-mode BL~Her stars 1--4~d), (ii) {\it I}-band luminosity and
$W_I$ Wesenheit index that placed the star up to 0.3~mag above or below the
red dashed lines shown in Figure~\ref{fig1} (which is comparable to
maximum dispersions of the PL relations obeyed by other types of classical
pulsators), (iii) {\it I}-band peak-to-peak amplitude larger than 0.15 mag
(to avoid low-amplitude non-radial pulsators and other types of
variable stars). Using these criteria we selected several thousand
variable stars in both Magellanic Clouds and carefully examined their light
curves.

\begin{figure*}
\plotone{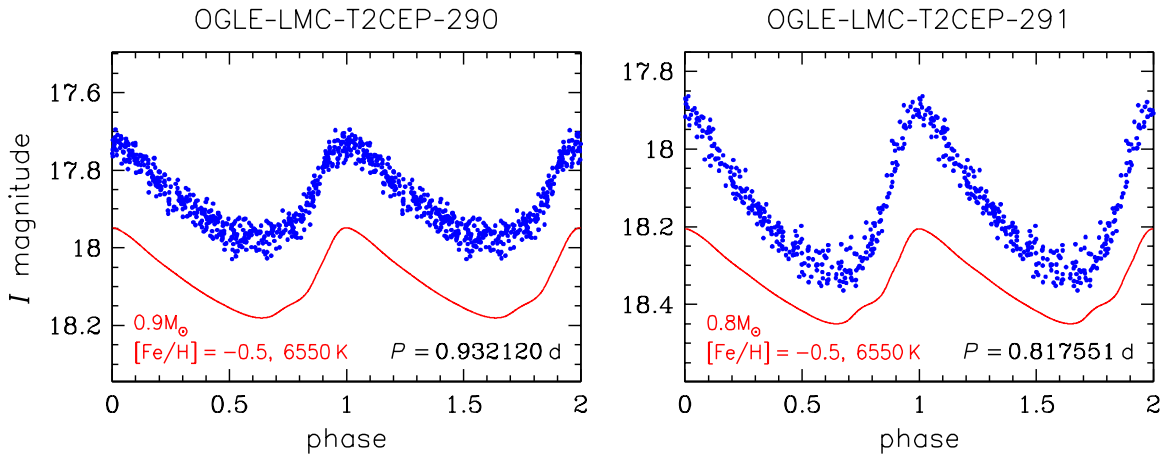}
\caption{{\it I}-band light curves of the BL~Her stars pulsating in
  the fist-overtone. Solid red lines shifted by 0.2\,mag correspond to
  the best matching non-linear pulsation models for OGLE-LMC-T2CEP-290
  and OGLE-LMC-T2CEP-291 (see Section~\ref{sec:models}). Physical
  parameters of the models -- mass, metallicity, and $T_\mathrm{eff}$
  -- are given in each panel.\label{fig2}}
\end{figure*}

As a result, we chose two pulsating variables which we believe
are the first-overtone BL~Her stars. Both objects are located in the
LMC. Their positions on the PW and PL diagrams are shown by red
circles in Figure~\ref{fig1}. In Figure~\ref{fig2}, we present their
{\it I}-band light curves phased with periods 0.932120~d and 0.817551~d,
respectively. No additional periodicities have been found in these light
curves.

These two stars have been added to the OGLE Collection of Type II
Cepheids in the LMC \citep{soszynski2018}. Their time-series OGLE-IV
photometry and finding charts are available through the FTP site:
\begin{center}
\url{ftp://ftp.astrouw.edu.pl/ogle/ogle4/OCVS/lmc/t2cep/}\\
\end{center}
Both objects received unique identifiers which follow the scheme
proposed in the OGLE Collection: OGLE-LMC-T2CEP-290 and
OGLE-LMC-T2CEP-291. Table~\ref{tab:table1} contains their basic
parameters: identifiers, J2000.0 equatorial coordinates, pulsation
periods, mean magnitudes in the {\it I} and {\it V} bands, and
{\it I}-band amplitudes.

\begin{deluxetable*}{ccccccc}
\tablecaption{First-overtone BL~Her stars in the LMC \label{tab:table1}}
\tablehead{
\colhead{Identifier} & \colhead{R.A.} & \colhead{Dec.} & \colhead{Period} & \colhead{$\langle{I}\rangle$} & \colhead{$\langle{V}\rangle$} & \colhead{$A(I)$} \\
\colhead{} & \colhead{[J2000.0]} & \colhead{[J2000.0]} & \colhead{[d]} & \colhead{[mag]} & \colhead{[mag]} & \colhead{[mag]}
}
\startdata
OGLE-LMC-T2CEP-290 & 04:51:11.61 & $-$69:00:33.0 & 0.9321202 & 17.869 & 18.520 & 0.238 \\
OGLE-LMC-T2CEP-291 & 05:42:38.85 & $-$61:59:44.3 & 0.8175507 & 18.135 & 18.647 & 0.417 \\
\enddata
\end{deluxetable*}

\begin{figure*}
\plotone{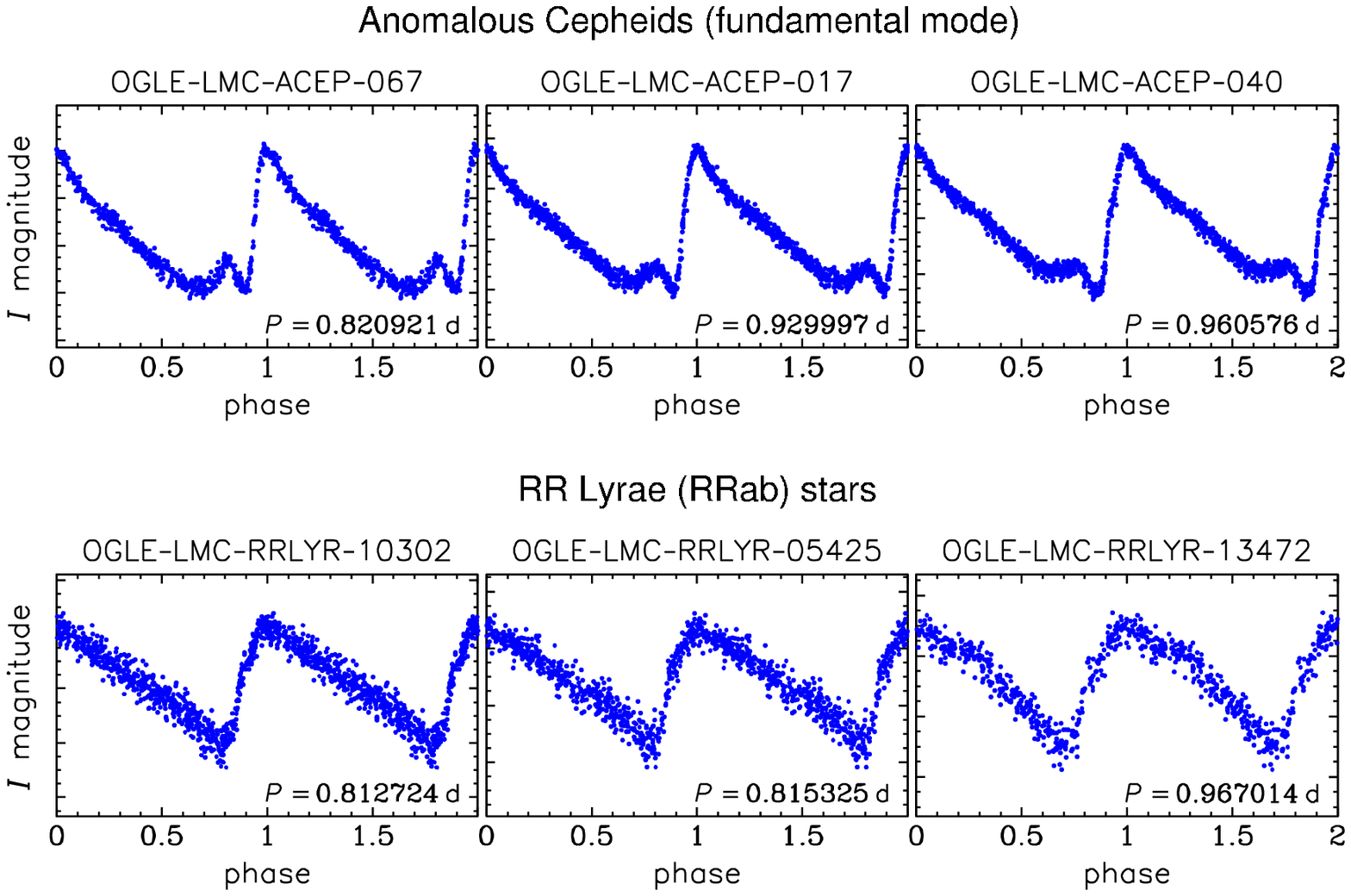}
\caption{Example {\it I}-band light curves of fundamental-mode
  anomalous Cepheids (upper panels) and RRab stars (lower panels) in
  the LMC. These stars have pulsation periods similar to the periods
  of the two first-overtone BL~Her stars (Figure~\ref{fig2}), but they
  have very different light curve shapes, in particular close to the
  minimum light.\label{fig3}}
\end{figure*}

\section{Discussion} \label{sec:dis}

\subsection{Shapes of the Light Curves} \label{sec:lcs}

In this section, we will provide evidence that OGLE-LMC-T2CEP-290 and
OGLE-LMC-T2CEP-291 are indeed stars pulsating in the
first-overtone. First, based on the light curve morphology
(Figure~\ref{fig2}), there is no doubt that both objects are radially
pulsating stars. Theoretically, spotted rotating stars may produce
light curves that mimic any type of periodic variables, including
pulsators, but such light curves are unstable and change shapes on a
time scale of months as the spot coverage changes on the stellar
surface. Meanwhile, the light curves of our objects have been stable
for years: at least from 2001 for OGLE-LMC-T2CEP-290 (which has been
monitored by the OGLE-III and OGLE-IV surveys) and from 2010 for
OGLE-LMC-T2CEP-291 (observed during the OGLE-IV
project). Characteristic shapes of the light curves indicate that both
stars are pulsating variables.

Second, the position of our objects on the PW and PL diagrams
(Figure~\ref{fig1}) is close to the expected relation for the overtone
BL~Her stars. Although OGLE-LMC-T2CEP-290 is slightly brighter, while
OGLE-LMC-T2CEP-291 slightly fainter than the theoretical relation for
the overtone type II Cepheids, but their dispersion around this
relation is typical for other classical pulsators in the LMC.

As one can see in Figure~\ref{fig1}, the PW and PL relations for the
first-overtone BL~Her stars partly overlap the relations for the
fundamental-mode RR Lyr variables (RRab stars) and fundamental-mode
anomalous Cepheids. For this reason, both our candidates for overtone
pulsators have already been included in the OGLE Collection of
Variable Stars and classified as a fundamental-mode anomalous Cepheid
(OGLE-LMC-T2CEP-290) and as an RRab star (OGLE-LMC-T2CEP-291).
However, a closer look at the light curves of our objects indicates
that they are not fundamental-mode pulsators.

\begin{figure*}
\plotone{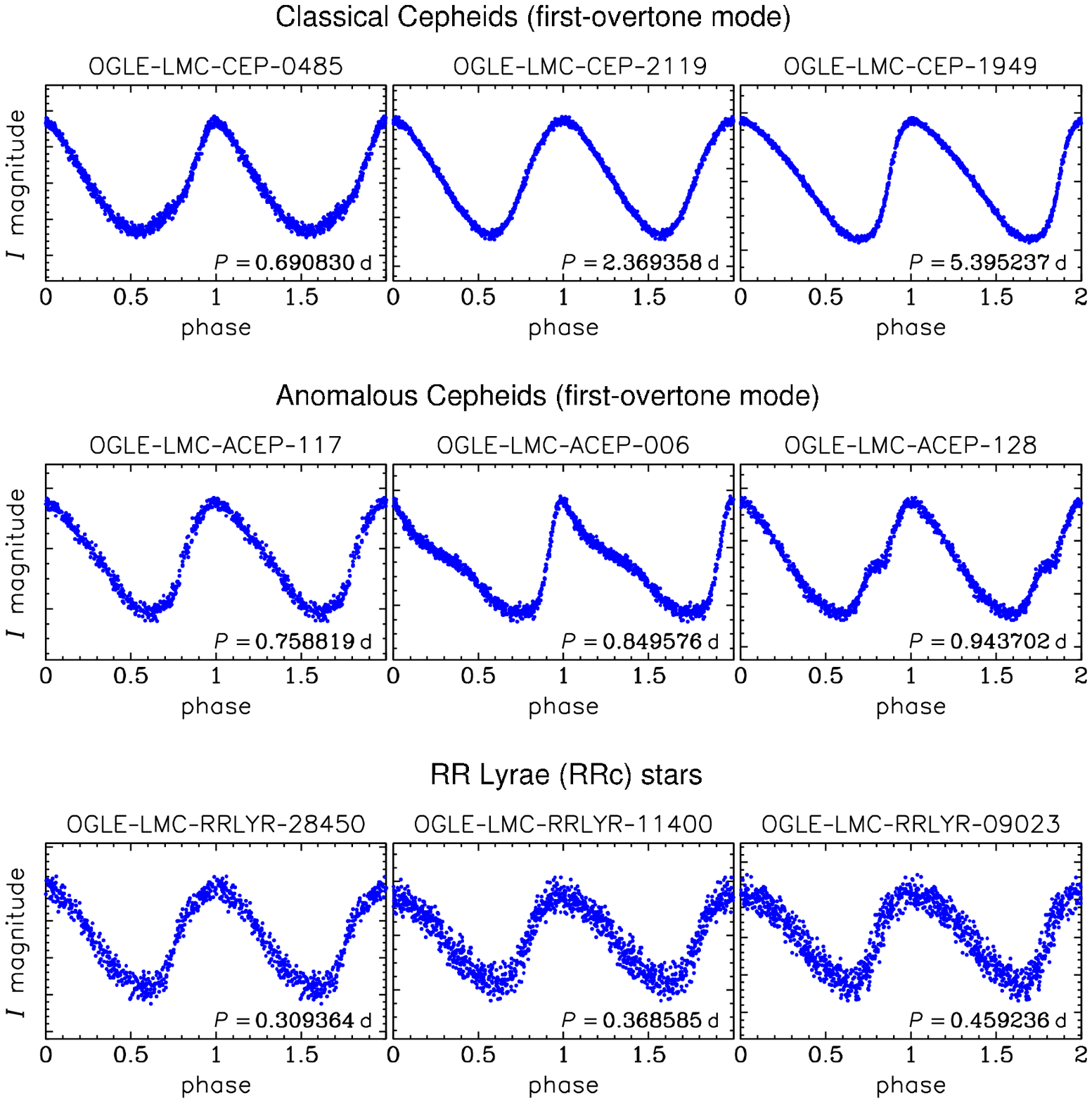}
\caption{Example {\it I}-band light curves of first-overtone classical
  pulsators in the LMC: classical Cepheids (upper panels), anomalous
  Cepheids (middle panels), and RR Lyr stars (lower panels). Note the
  round minima in all these light curves, just like in the
  first-overtone BL~Her stars (Figure~\ref{fig2}).\label{fig4}}
\end{figure*}

In Figure~\ref{fig3}, we present three light curves of typical
fundamental-mode anomalous Cepheids and three light curves of RRab
stars from the LMC. Here, we selected variables with periods similar
to the periods of our candidates for the first-overtone BL~Her stars
(around 0.82 or 0.93~days). The difference between the light curves in
Figures~\ref{fig2} and \ref{fig3} is visible to the naked eye. In
particular, we draw the reader's attention to the shape of the light
curves close to the minimum brightness. Anomalous Cepheids with these
periods exhibit a small bump just before the sharp minimum. RRab stars
also have sharp minima, while both our candidates for the
first-overtone BL~Her stars show round minima.

Figure~\ref{fig4} displays a sample of {\it I}-band light curves
belonging to the first-overtone classical pulsators in the LMC:
classical Cepheids, anomalous Cepheids and RR Lyr stars (RRc
variables). We selected typical overtone variables covering a wide
range of pulsation periods. The morphology of the light curves varies
from star to star, but one feature is common for all stars with the
overtone modes excited -- they have round minima of their light
curves. This is a very strong argument that OGLE-LMC-T2CEP-290 and
OGLE-LMC-T2CEP-291 belong to the same (first-overtone) class of
pulsating stars.

\begin{figure*}
\plotone{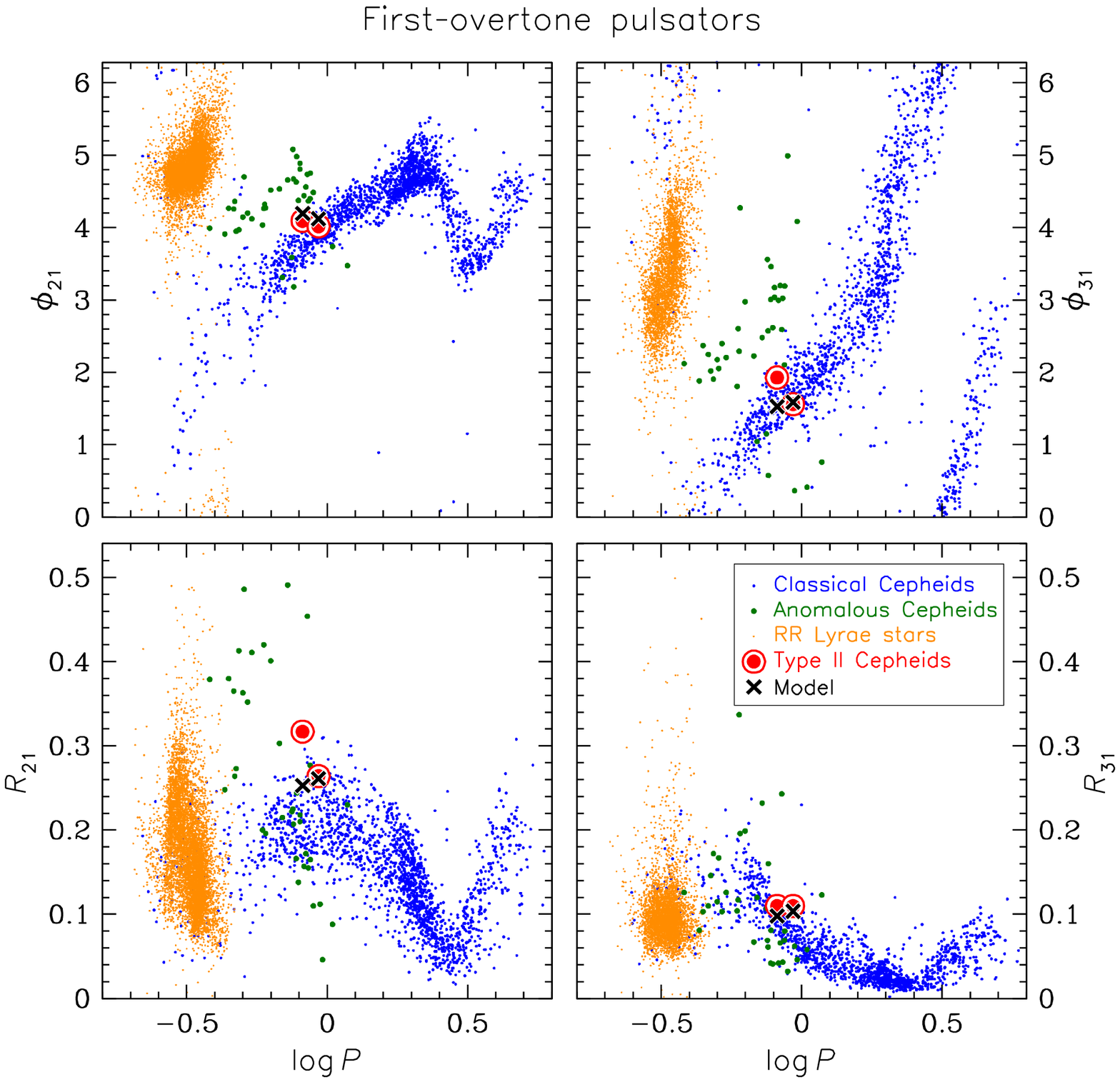}
\caption{Fourier parameters $\phi_{21}$, $\phi_{31}$, $R_{21}$, and
  $R_{31}$ as a function of $\log{P}$ for first-overtone classical
  pulsators in the LMC. Different symbols indicate the same types of
  stars as in Figure~\ref{fig1}. In addition, black crosses correspond
  to the Fourier parameters of the best matching non-linear pulsation
  models for OGLE-LMC-T2CEP-290 and OGLE-LMC-T2CEP-291 (see
  Section~\ref{sec:models}).\label{fig5}}
\end{figure*}

\begin{figure*}
\plotone{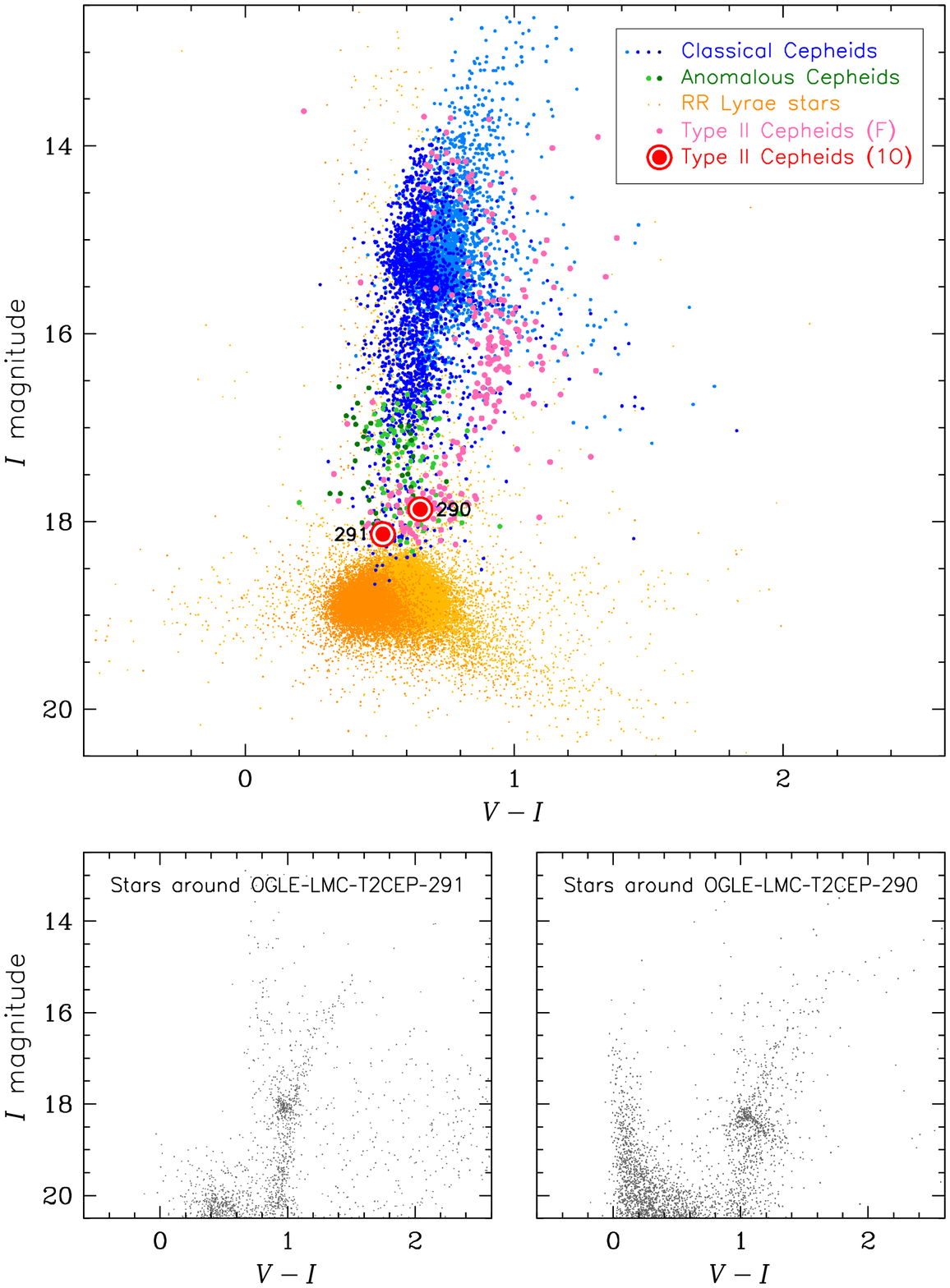}
\caption{Apparent $(V-I)$ vs. $I$ color--magnitude diagrams for stars
  in the LMC. Upper panel shows classical pulsators from the OGLE
  Collection of Variable Stars. Different symbols indicate the same
  types of stars as in Figure~\ref{fig1}. Lighter and darker colors
  indicate fundamental-mode and overtone-mode pulsators, respectively.
  Lower panels show all stars in the vicinities of the two first-overtone
  BL~Her stars: up to 10~arcmin from OGLE-LMC-T2CEP-291 (left panel), and
  up to 5~arcmin from OGLE-LMC-T2CEP-290 (right panel).\label{fig6}}
\end{figure*}

The coefficients derived from the Fourier decomposition of light
curves provide a quantitative description of their morphology. We
fitted the cosine Fourier series to the {\it I}-band light curves of
our variables and derived the $\phi_{21}$, $\phi_{31}$, $R_{21}$, and
$R_{31}$ coefficients \citep{simon1981}. Figure~\ref{fig5} compares
these parameters for first-overtone classical pulsators in the
LMC. The light curves of our candidates seem to be the most similar to
the first-overtone classical Cepheids, although at the considered
period range, Fourier parameters for anomalous Cepheids and classical
Cepheids partially overlap. This similarity confirms that
OGLE-LMC-T2CEP-290 and OGLE-LMC-T2CEP-291 are the first overtone pulsating
stars, although they do not fit perfectly in each diagram which is expected
since they are members of a different class of pulsators than classical and
anomalous Cepheids. Our candidates have smaller Fourier phases $\phi_{21}$
and $\phi_{31}$ (upper panels of Figure~\ref{fig5}) than typical anomalous
Cepheids with the same periods and slightly larger amplitude ratios
$R_{21}$ and $R_{31}$ (lower panels of Figure~\ref{fig5}) than most of
classical Cepheids. The Fourier parameter diagrams nicely confirm the
first-overtone pulsation of OGLE-LMC-T2CEP-290 and OGLE-LMC-T2CEP-291 and,
together with the PL and PW diagrams, discriminate between variability
classes.

\subsection{Position in the Color--Magnitude Diagram} \label{sec:cmd}

Another indicator of the pulsation mode is the position of a star in
the color--magnitude diagram. It is known that the overtone pulsators
occupy the blue side of the instability strip. Their $(V-I)$ colors
are on average 0.10--0.15~mag smaller than for the fundamental-mode
Cepheids and RR Lyr stars, although there is an overlap between
different pulsation modes. Additionally, the apparent color indices of
individual objects may be strongly affected by the interstellar
reddening or blending by an unresolved companion, which limits the
applicability of the color criterion for determining the pulsation
modes.

In the upper panel of Figure~\ref{fig6}, we present the $(V-I)$
vs. $I$ color--magnitude diagram for classical pulsators in the
LMC. As in Figure~\ref{fig1}, darker colors indicate overtone
variables. OGLE-LMC-T2CEP-291 has an apparent color index
$(V-I)=0.51$~mag, which agrees with typical colors of other overtone
classical pulsators and is bluer than most of the fundamental-mode
BL~Her stars. This is another direct proof that OGLE-LMC-T2CEP-291
pulsates in the first-overtone mode.

However, the color index of OGLE-LMC-T2CEP-290 (0.65~mag) places this
star closer to the fundamental-mode pulsators. This fact can be
explained by a high interstellar reddening toward this object. Lower
panels of Figure~\ref{fig6} present color--magnitude diagrams for all
stars (included in the OGLE-IV database) within a radius of 10~arcmin
(for OGLE-LMC-T2CEP-291) and 5~arcmin (for OGLE-LMC-T2CEP-290) around
both candidates for the overtone type II Cepheids. The surroundings of
OGLE-LMC-T2CEP-291 (lower left panel of Figure~\ref{fig6}) are
characterized by well-defined red clump with the mean apparent color
index of $(V-I)=0.97$~mag. On the other hand, in the vicinity of
OGLE-LMC-T2CEP-290 (lower right panel), the red clump is significantly
elongated toward redder and fainter stars, which indicates high and
differential reddening toward these regions. The mean apparent $(V-I)
$ color index of the red clump stars in this area cannot be precisely
determined -- we can only specify that it is included in the range
from 1.01 to 1.22~mag. It translates into a range from 0.04 to
0.25~mag of higher reddening toward OGLE-LMC-T2CEP-290 than to
OGLE-LMC-T2CEP-291, which fully explains the redder color of the
former object.

\subsection{Spatial Location} \label{sec:space}

\begin{figure*}[t]
\plotone{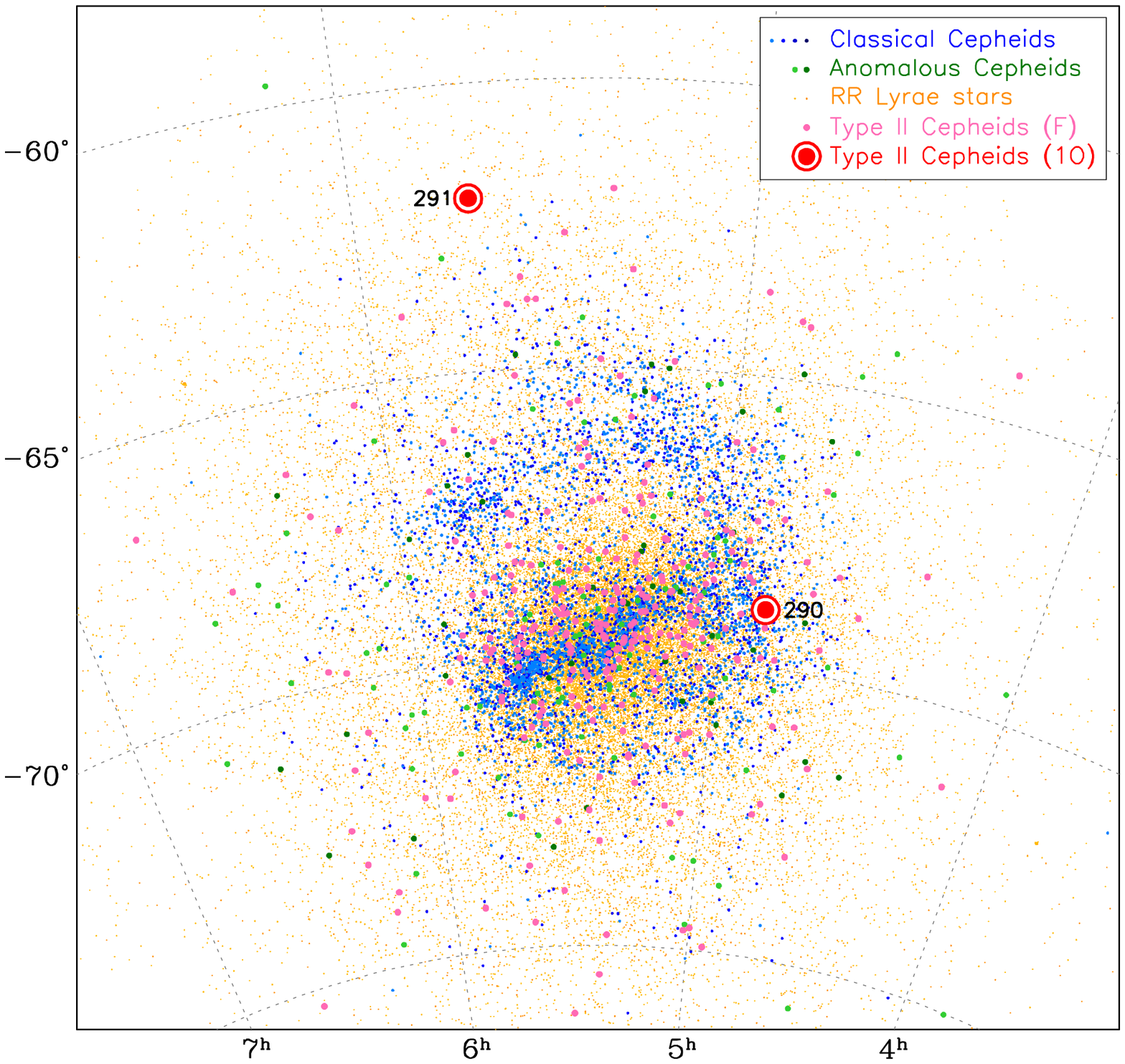}
\caption{Sky map of the LMC with classical pulsators from the OGLE
  Collection of Variable Stars. Different symbols indicate the same
  types of stars as in Figure~\ref{fig1}.\label{fig7}}
\end{figure*}

Figure~\ref{fig7} displays the sky map of the LMC with classical
pulsators marked with the same symbols as in
Figure~\ref{fig1}. OGLE-LMC-T2CEP-290 is located closer to the LMC
center, on the western edge of the LMC bar, while OGLE-LMC-T2CEP-291
is located in the northern outskirts of the LMC, where classical
Cepheids are absent, but the old stellar population (RR Lyr stars) is
still common. This confirms that at least OGLE-LMC-T2CEP-291 belongs
to the old population.

\begin{figure*}
\plotone{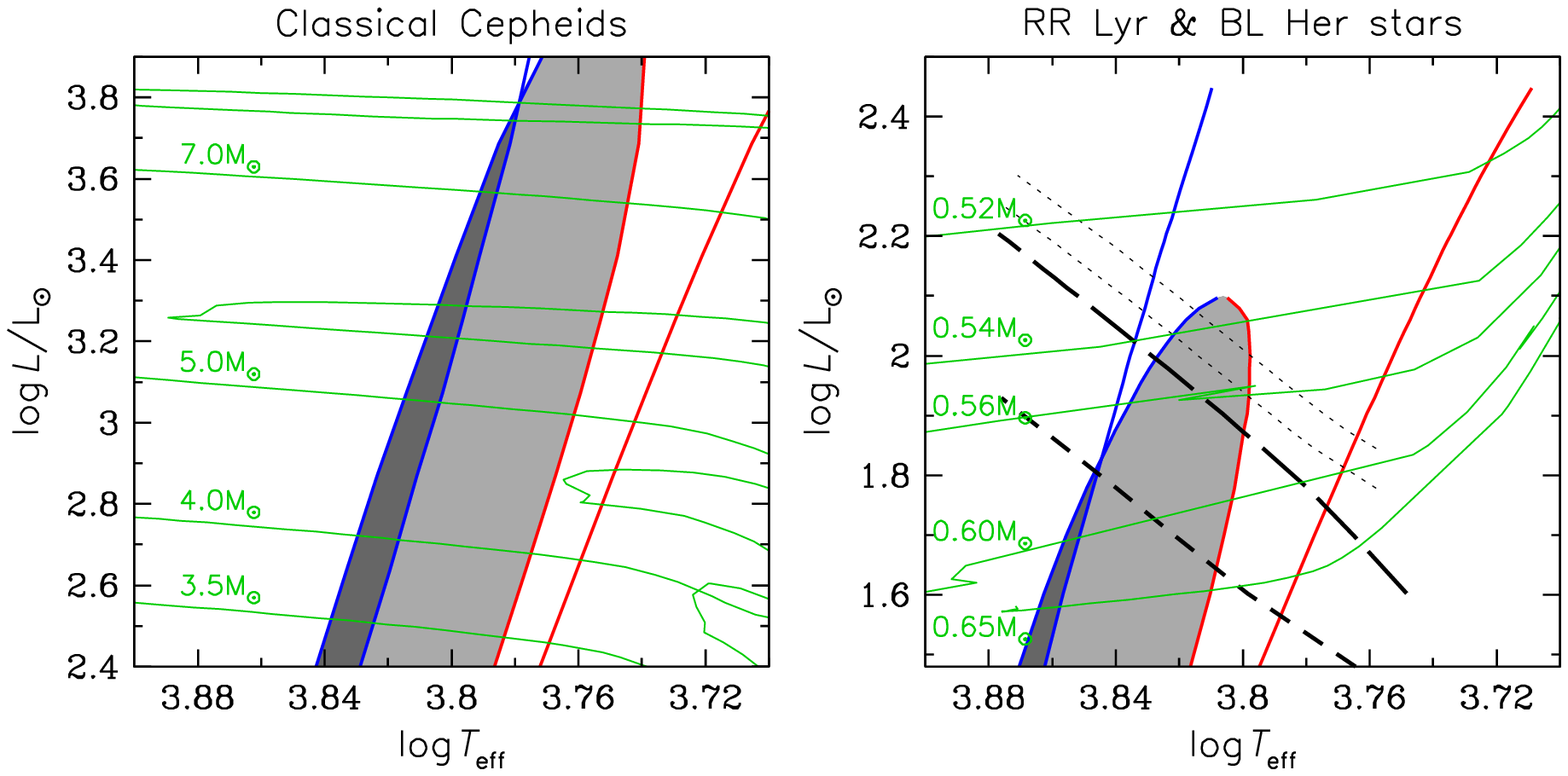}
\caption{Qualitative view on pulsation and evolutionary scenarios of
  classical Cepheids (left panel) and RR~Lyr stars and BL~Her
  variables (right panel). Solid blue and red lines show the blue and
  red edges of the classical instability strip. Instability strip for
  the first-overtone mode is gray-shaded. In the dark-gray-shaded area only
  the first-overtone is linearly unstable, while in the light-gray-shaded
  area both first-overtone and fundamental modes are linearly
  unstable. Example evolutionary tracks are plotted in each panel with
  solid green lines and labeled with corresponding masses. In the right
  panel thick long-dashed line is placed at constant fundamental-mode
  period of 1~d and separates RR~Lyr and BL~Her domains. Thick short-dashed
  line is placed at constant first-overtone period of 0.44~d. Observations
  indicate that majority of RRc and RRd stars have shorter periods. Two
  thin dotted lines are placed at constant first-overtone periods
  corresponding to pulsation periods of OGLE-LMC-T2CEP-290 and
  OGLE-LMC-T2CEP-291.}
\label{fig:hr_overview}
\end{figure*}

What is the distance of our candidates for the first-overtone BL~Her
stars? So far we have assumed that both objects are located in the LMC
and their position on the apparent PL diagram (Figure~\ref{fig1})
reflects their position in the absolute PL plane. However, there is a
possibility that we deal with first-overtone classical or anomalous
Cepheids located behind the LMC. Taking into account the position of
both stars in the extinction-free PW diagram, we may estimate that if
they are the first-overtone classical Cepheids, they should be located
about twice as far as the center of the LMC. It seems very unlikely
that stars belonging to the young stellar population (classical
Cepheids) exist in the intergalactic space. In turn, assuming that
OGLE-LMC-T2CEP-290 and OGLE-LMC-T2CEP-291 are the first-overtone
anomalous Cepheids, they should be located 50-60\% farther than the
LMC. Taking into account that the light curves of our objects
(quantified by the Fourier coefficients; Figure~\ref{fig5}) are
different than light curves of typical first-overtone anomalous
Cepheids with the same periods, such a possibility also seems
unlikely.

\subsection{First-Overtone BL~Her Stars from Theoretical Perspective}\label{sec:models}

While single-periodic first-overtone pulsation is common among
classical Cepheids and RR~Lyr stars, it has not been detected so far
in any type II Cepheid. Here we explain why first-overtone pulsation
for type II Cepheids must be very rare. In
Figure~\ref{fig:hr_overview}, we show the pulsation and evolutionary
scenarios for classical Cepheids (left panel) and for RR~Lyr stars and
the shortest-period type II Cepheids -- BL~Her stars (right
panel). Instability strips for the fundamental mode (F mode; solid
blue and solid red lines) and for the first-overtone mode (1O mode;
gray-shaded area) were computed using linear convective codes of
\cite{smolec2008a}. In all computations presented in this paper
parameters of the turbulent convection model are the same as we used
in the successful modeling of a period-doubled BL~Her star
\citep{smolec2012}. For classical Cepheids, the model grid was
constructed assuming mass--luminosity relation inferred from
evolutionary models of \cite{anderson2016} for ${\rm [Fe/H]}=-0.5$. In
the right-hand panel (RR~Lyr \& BL~Her stars) we assumed a constant
mass of $0.60\MS$ and ${\rm [Fe/H]}=-1.0$. Evolutionary tracks for
classical Cepheids are from \cite{bertelli2009} ($Z=0.008$) and
evolutionary tracks for horizontal-branch stars are from
\cite{dotter2008} (${\rm [Fe/H]}=-1.0$).

We note that model calculations are sensitive e.g.\ to the physical
parameters, treatment of the convection, or details of the evolutionary
modeling. For example, different extent of overshooting during main
sequence evolution can shift the evolutionary tracks for Cepheids in
luminosity and affect the extent of the blue loops. By adopting different
parameters for turbulent convection model in pulsation calculations, the
location of the instability strips may, e.g. slightly shift in effective
temperature or in luminosity, but the overall shape and features of the
instability strip, like limited extent of 1O instability strip in
luminosity, will be the same. Here we just discuss the qualitative
pulsation and evolutionary picture and these details are inessential for
our discussion.

For classical Cepheids (left panel of Figure~\ref{fig:hr_overview}) we
can identify a domain in which only 1O is linearly unstable (dark-gray
shaded area). If a star moves toward lower effective temperatures
(first and third crossing of the instability strip) it first enters
the domain in which only 1O is linearly unstable and single-periodic
1O pulsation is the only possibility. Then, it enters the domain in
which both modes, F and 1O, are linearly unstable (light-gray shaded
area), and finally a domain in which F mode pulsation is the only
possibility. What happens in the domain in which both modes are
linearly unstable is dictated by non-linear mode selection \citep[see, 
e.g.][]{smolec2008b}. The star that enters this domain from the hot
side may continue the single-periodic 1O pulsation until it switches
to F mode single-periodic pulsation somewhere within the
domain. Similarly, the star that enters this domain from the cool side
may continue the F mode pulsation, until it switches to 1O single-mode
pulsation somewhere within the domain. In a part of the HR diagram
within light-gray shaded area the so-called {\it either-or} scenario
is possible: the star may pulsate either in 1O mode or in F mode
depending on the direction of evolution. The other possible scenario
is double-mode pulsation: simultaneous pulsation in both F and 1O
modes. Only with non-linear calculations this mode selection problem
can be addressed. It is, however, clear that over significant part of
the HR diagram single-periodic 1O pulsation in classical Cepheids is
possible. First-overtone classical Cepheids, as well as double-mode
F+1O classical Cepheids are common.

For RR~Lyr stars (right panel of Figure~\ref{fig:hr_overview}; stars
with fundamental-mode periods below $1$\,d, i.e. below the thick
long-dash black line) the scenario is qualitatively the same as just
outlined for classical Cepheids. We can identify the domain in which
1O pulsation is the only possibility (hottest pulsators; dark-gray
shaded area) and a domain in which both 1O and F modes are linearly
unstable (light-gray shaded area), in which either single-periodic 1O,
or single-periodic F mode pulsation, as well as double mode F+1O
pulsation are possible. Indeed, both single-periodic 1O pulsators (RRc
stars) and double-mode pulsators (RRd stars) are common (as well as
single-periodic F mode pulsators, RRab stars).

At this point it is essential to point that there is a maximum period
for 1O RR~Lyr pulsators. Considering the most numerous OGLE collection
of RR~Lyr variables \citep{soszynski2014,soszynski2016} the longest
possible 1O periods are approximately $0.539$, $0.495$ and $0.487$\,d
for RRc stars of the Galactic bulge, LMC, and SMC, respectively and
$0.451$, $0.521$ and $0.524$\,d for RRd stars of the same stellar
systems. These numbers are all quite similar; differences often rely
on a single record-holder star. In 99\,\% of the above considered
stars from the OGLE collection, 1O period is shorter than $0.44$\,d --
a value marked with thick short-dashed line in
Figure~\ref{fig:hr_overview}. Observations clearly indicate that above
this line nearly all RR~Lyr stars pulsate in F mode only, despite 1O
can still be linearly unstable there. We note that above the discussed
line, 1O domain is essentially engulfed by the domain in which F mode
is unstable. Consequently, before entering the domain in which both F
and 1O modes are linearly unstable, a star was necessarily
single-periodic F mode pulsator.

\begin{figure*}
\plotone{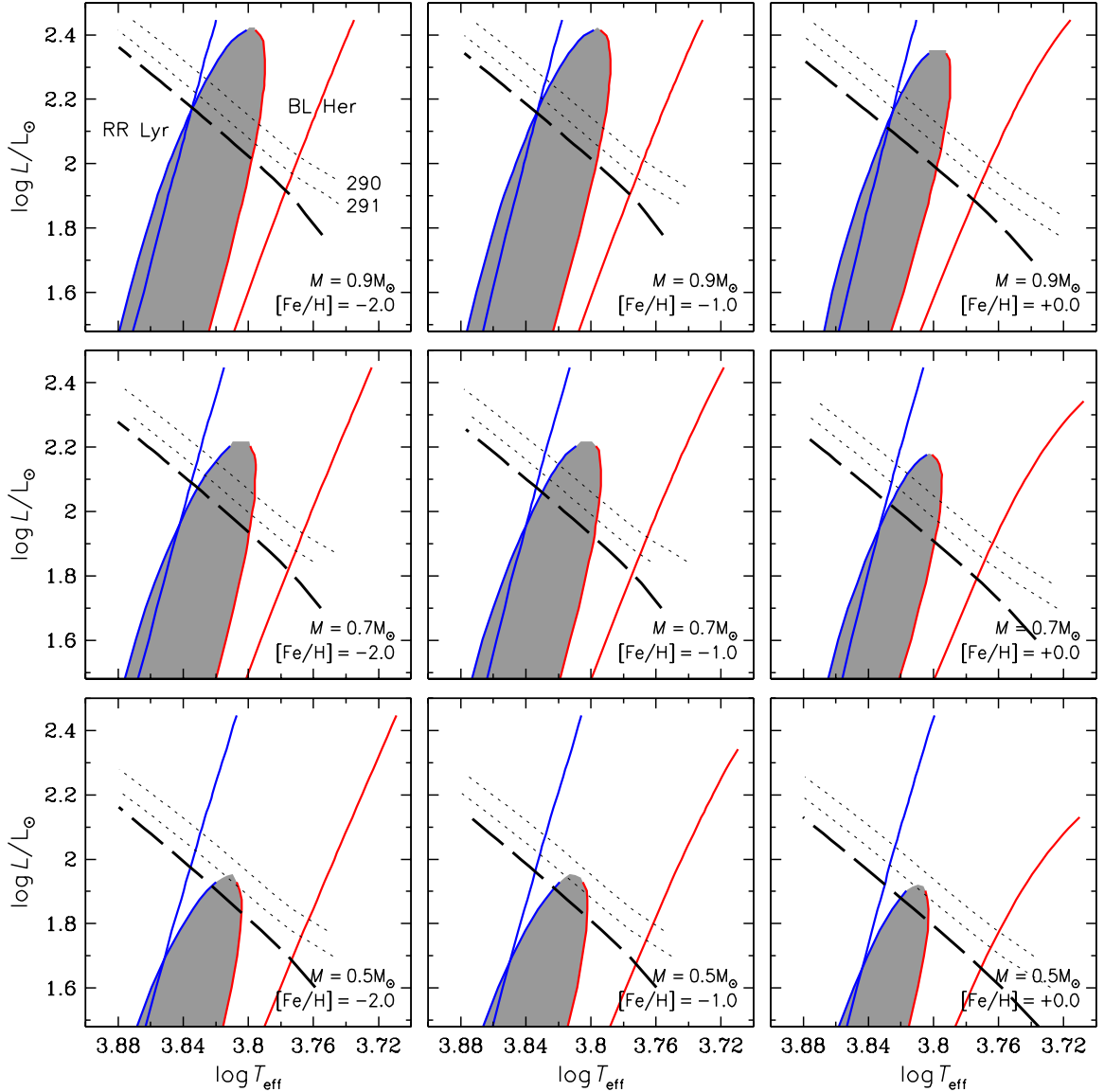}
\caption{HR diagrams with linear instability strips for selected
  models of $0.9\MS$ (top row), $0.7\MS$ (middle row) and $0.5\MS$
  (bottom row). Metallicity is constant in each column: $-2.0$ (left
  column), $-1.0$ (middle column) and $+0.0$ (left column). Solid blue
  and red lines show the blue and red edges of the classical
  instability strip. Instability strip for the first-overtone mode is
  gray-shaded. Thick long-dashed line is placed at constant
  fundamental-mode period of 1\,d and separates RR~Lyr and BL~Her
  domains. Two thin dotted lines are placed at constant first-overtone
  periods corresponding to pulsation periods of OGLE-LMC-T2CEP-290 and
  OGLE-LMC-T2CEP-291, as indicated in the top left panel.}
\label{fig:hr_grid}
\end{figure*}

The lack of long-period RRc/RRd stars indicates that non-linear mode
selection favors the F-mode-only pulsation also in the domain in which
both modes are linearly unstable. A star that was pulsating in the F
mode continues the F mode pulsation also after entering the domain in
which 1O mode becomes linearly unstable. RR~Lyr stars and BL~Her
variables are close siblings \citep[e.g.][]{iwanek2018}. BL~Her stars
are just slightly less massive, start the HB evolution at higher
effective temperatures and cross the instability strip at larger
luminosities. The instability strip displayed in
Figure~\ref{fig:hr_overview} was computed for $M=0.6\MS$, however for
lower masses, the 1O instability strip shrinks and extends to lower
luminosities (see Figure~\ref{fig:hr_grid}, bottom panels) so
long-period single-periodic 1O pulsation is even more excluded for
BL~Her stars.

Still, as we determined on observational basis, OGLE-LMC-T2CEP-290 and
OGLE-LMC-T2CEP-291 with much longer periods are single-periodic 1O
pulsators. To understand the nature of these two stars we have
computed a small model survey covering a much larger range of masses
than is typically considered for BL~Her-type stars. Masses in the grid
cover $0.5-0.9\MS$ range with $0.05\MS$ step. Their metallicities are
in the $-2.5,\ldots,+0.0$ range with 0.5~dex step.

Instability strips for selected models are plotted in the HR diagrams
in Figure~\ref{fig:hr_grid}, along with lines of constant period equal
to the periods of OGLE-LMC-T2CEP-290 and OGLE-LMC-T2CEP-291. In
addition, thick long-dashed line is placed at constant
fundamental-mode period equal to 1\,d and thus separates the RR~Lyr
and BL~Her classes. We find that the lower the mass the shorter the
extension of 1O instability strip in luminosity. In particular, for
$M=0.5\MS$, there are no models for OGLE-LMC-T2CEP-290 in which the
first overtone is linearly unstable. The line of constant period
corresponding to pulsation period of OGLE-LMC-T2CEP-291 crosses the 1O
instability domain at its tip. For larger masses, 1O is linearly
unstable for both stars over a significant range of effective
temperatures. We also note that along the discussed lines of constant
1O period, 1O instability domains are always engulfed by F mode
instability domain. Systematics with metallicity is difficult to spot
in Figure~\ref{fig:hr_grid}. We note that with increasing metallicity
the instability strips shift toward lower effective
temperatures. Also, the larger the metallicity the narrower the 1O
instability strip. Based on linear models we may expect that the
larger the mass the higher the chances for 1O pulsation for
OGLE-LMC-T2CEP-290 and OGLE-LMC-T2CEP-291. However, only with
non-linear calculations one may study for which physical parameters
stable large amplitude 1O pulsation is possible.

Non-linear models were computed using the codes of
\cite{smolec2008a}. Models were computed for the same grid of masses
and metallicities as in the linear model survey discussed above. For a
given mass and metallicity, the models were computed along the lines
of constant 1O periods, corresponding to the periods of
OGLE-LMC-T2CEP-290 and OGLE-LMC-T2CEP-291, with a $50$\,K-step in
effective temperature. These models were computed only across the
domain in which the first-overtone is linearly unstable. As it is
clear from Figure~\ref{fig:hr_grid}, the fundamental mode is
necessarily unstable as well then. The static models were perturbed
with the first-overtone scaled velocity eigenfunction calculated for
each model during linear analysis, and time integration was conducted
for at least $6\,000$ pulsation cycles.

\begin{deluxetable*}{cccc}
\tablecaption{Physical parameters of non-linear models computed for
  OGLE-LMC-T2CEP-290 and OGLE-LMC-T2CEP-291 that converged to full
  amplitude single-periodic first-overtone
  pulsation. \label{tab:table2}}
\tablehead{
ID & mass [\MS]& ${\rm [Fe/H]}$ & $T_{\rm eff}$ [K]
}
\startdata
      OGLE-LMC-T2CEP-290 & 0.85  & $-0.5$ & 6450--6500\\
                &       & $-1.0$ & 6350--6500\\
                &       & $-1.5$ & 6400--6450\\
                & 0.90  & $-0.5$ & 6350--6550\\
                &       & $-1.0$ & 6300--6500\\
      \hline
      OGLE-LMC-T2CEP-291 & 0.75  & $-0.5$ & 6450--6550\\
                &       & $-1.0$ & 6400--6550\\
                & 0.80  & $-0.5$ & 6350--6600\\
                &       & $-1.0$ & 6350--6550\\
                & 0.85  & $+0.0$ & 6400--6550\\
                &       & $-0.5$ & 6350--6600\\
                &       & $-1.0$ & 6400--6500\\
                & 0.90  & $+0.0$ & 6350--6600\\
                &       & $-0.5$ & 6350--6550\\
\hline
\enddata
\end{deluxetable*}

During the model integration, after initial transient phase in which
first-overtone pulsation was present, the majority of models switched
into single-mode fundamental-mode pulsation. It was the case for all
models with masses below $0.75\MS$. It well agrees with the
considerations presented above: with parameters expected for typical
BL~Her stars single-mode first-overtone pulsation is not possible. As
mass is increased in the models we observe that the switching from the
first-overtone pulsation to fundamental-mode pulsation occurs later
during the integration. For models of OGLE-LMC-T2CEP-291 (with shorter
pulsation period, $P\approx0.818$\,d), single-mode first-overtone
pulsation becomes possible for $M\ge0.75\MS$. For slightly more
luminous and longer-period models of OGLE-LMC-T2CEP-290
($P\approx0.932$\,d) first-overtone pulsation becomes possible only
for $M\ge0.85\MS$. In Table~\ref{tab:table2} we provide the physical
parameters of the models that converged for single-mode 1O
pulsation. We find that 1O pulsation is not possible for the lowest
considered metallicities, i.e. for $-2.5$ and $-2.0$. For
${\rm [Fe/H]}=-1.5$ only a single model converged to 1O pulsation
($M=0.85\MS$). 1O pulsation is possible for ${\rm [Fe/H]}\ge-1.0$. For
the highest considered mass 1O models are present also for solar
metallicity. The higher the mass, the wider the domain of 1O
pulsation.

A rather large domains over which models converged to full amplitude
1O pulsation may indicate that such pulsation should be commonly
observed. We recall however, that the way we initialized model
integration -- low-amplitude perturbation with scaled velocity
eigenvector of 1O mode -- favors 1O pulsation. As it is clear from
Figure~\ref{fig:hr_grid}, the star that enters the domain in which 1O
mode is linearly unstable is already a large amplitude F mode pulsator
and may continue the F mode pulsation in the domain in which both F
and 1O modes are unstable. In principle, using non-linear calculation
one can study the stability of full-amplitude F mode pulsation with
respect to perturbation in 1O mode. This is, however, very
time-consuming as it requires several integrations of the same model
with different initial conditions \citep[see, e.g.][]{smolec2008b}.
Such analysis is beyond the scope of the present paper, but it is
planned.

The computed bolometric light curves were transformed to the $I$-band
via static \cite{castelli2004} model atmospheres. In
Figure~\ref{fig2}, we compare the observed light curves with the best
matching model, i.e. the model in which relative differences of
low-order Fourier coefficients between the model and observed light
curve are the lowest. Physical parameters of these models are given in
the panels. We note that the pulsation amplitude in the best model for
OGLE-LMC-T2CEP-291 is nearly half of that observed. We comment on this
difference more in the following. We note that the light curves of all
computed first-overtone-models are all very similar to each other and
that plotted in Figure~\ref{fig2} may be regarded as
representative. The match between the model and observed light curves
is reasonable. In the models, the change of inclination on the
ascending branch, after the minimum brightness is quite pronounced,
while in the observed light curves it is well visible only for
OGLE-LMC-T2CEP-290. To compare the models and light curves in a more
quantitative way, in Figure~\ref{fig5} we have also plotted the
Fourier parameters of the best-matching models (large crosses). The
match for OGLE-LMC-T2CEP-290 is remarkable. For OGLE-LMC-T2CEP-291 the
match for Fourier phases is very good, while amplitude ratios are a
bit lower, which is expected, as pulsation amplitude is also lower
(Figure~\ref{fig2}).

Although the light variation of OGLE-LMC-T2CEP-290 and
OGLE-LMC-T2CEP-291 can be reproduced reasonably well, the
first-overtone non-linear models we have just discussed are
inconsistent with observations regarding their luminosity. When the
mass is increased, the luminosity must also be increased to match the
periods of the two modeled stars (see
Figure~\ref{fig:hr_grid}). Consequently, our models are both more
massive and more luminous than expected for BL~Her stars. In the PW
plane (Figure~\ref{fig1}), the models are placed at a position
characteristic for fundamental-mode anomalous Cepheids, in between
position typical for first-overtone anomalous Cepheids and position
expected for first-overtone BL~Her stars. The best-matching models for
OGLE-LMC-T2CEP-290 and OGLE-LMC-T2CEP-291 are by 0.36 and 0.58\,mag
more luminous than observed, respectively.

Our initial model survey indicates that indeed the light curve shape
observed for OGLE-LMC-T2CEP-290 and OGLE-LMC-T2CEP-291 is
characteristic for single-mode first-overtone pulsation. However,
single-mode first-overtone pulsation becomes possible only when mass
is significantly increased above the values expected for BL~Her
stars. It may indicate that our stars are a product of specific mass
transfer event during binary evolution. Binarity is often invoked as
plausible explanation for the properties of peculiar W~Vir stars or of
anomalous Cepheids \cite[e.g.][]{soszynski2017b}. In this scenario,
however, the stars should be more luminous than observed. Pulsation
modeling of OGLE-LMC-T2CEP-290 and OGLE-LMC-T2CEP-291 remains a puzzle
at the moment.

We stress that our models survey is very limited; a thorough analysis
that may help to resolve the raised issues, is planned. In particular,
the following should be addressed in a more detailed model survey. (i)
Full mode selection analysis should be done. It requires integration
of each model with various initial conditions and analysis of the
resulting hydrodynamic trajectories with amplitude equation
formalism. (ii) Other sets of convective parameters should be
investigated. In particular, models with decreased eddy-viscous
dissipation. Such models should have larger pulsation amplitudes and
hence should match the observations of OGLE-LMC-T2CEP-291 better. More
importantly, the extent of the instability domains and mode selection
strongly depend on eddy-viscous dissipation and other parameters that
enter the convection model used in the code. By varying these
parameters, we may hope for first-overtone pulsation at lower masses
and hence at lower luminosities. (iii) Extension of the model survey
to even higher masses, up to that considered for anomalous Cepheids
($\sim1.5\MS$), is desired.

\section{Conclusions}
We reported here the discovery of two candidates for single-mode
first-overtone type II Cepheids. OGLE-LMC-T2CEP-291 meets all
observational criteria that can be checked using the OGLE data:
positions in the PL and PW diagrams, light curve morphology, and
$(V-I)$ color index. OGLE-LMC-T2CEP-290 meets the first two
conditions, but it is too red as for the first-overtone pulsator,
however this fact can be explained by a higher interstellar reddening
toward this star. We believe that our detection is very reliable,
since the light curves of our two candidates have definitely different
shapes than fundamental-mode pulsators located in this region of the
PL diagram. Spectroscopic follow-up observations could bring new
arguments for or against our classification.

To our knowledge, OGLE-LMC-T2CEP-290 and OGLE-LMC-T2CEP-291 are the
first known type II Cepheids pulsating solely in the
first-overtone. \citet{mccollum1997} suggested that a bright BL~Her
star VY Pyx is an overtone pulsator, however a glance at the light
curve of this objects \citep{sanwal1991} reveals sharp minima, which
is a characteristic feature of the fundamental-mode pulsators. Also
the Hipparcos parallax place VY Pyx on the PL relation obeyed by the
fundamental-mode type II Cepheids \citep{feast2008}.

With the help of pulsation models we explained why first-overtone
pulsation of BL~Her stars must be very rare, or is hardly possible, at
least for masses typically assumed for these stars. For masses
$\sim0.5-0.6\MS$ the linear instability strip for the first-overtone
mode does not extend to sufficiently high luminosities to secure
first-overtone periods well above $0.7$\,d, in the range expected for
BL~Her variables. As mass is increased ($\sim0.6-0.7\MS$), this
restriction is no longer valid: linear first-overtone instability
strip extends to sufficiently high luminosities, at least to warrant
pulsation periods observed in OGLE-LMC-T2CEP-290 and
OGLE-LMC-T2CEP-291. However, the first-overtone instability strip is
entirely engulfed within fundamental-mode instability strip then. As
these are the masses expected also for RR~Lyr stars, observations,
namely the first-overtone periods observed in RRc and RRd stars
(always below $\approx0.5$\,d), point that non-linear mode selection
favors fundamental-mode pulsation in the high luminosity part of the
joint instability strip of the two radial modes. This was fully
confirmed with the help of non-linear pulsation calculations. The
model integration started with first-overtone perturbation always led
to fundamental-mode finite amplitude pulsation. Only for $M\ge0.75\MS$
the model integration finally ended in stable first-overtone
pulsation. We stress that the best matching model for
OGLE-LMC-T2CEP-290 ($0.9\MS$) matches the observed light curve
remarkably well. The best matching model for OGLE-LMC-T2CEP-291
($0.8\MS$) also well reproduces the observed light variation, only the
amplitude is about twice as low as observed. The successful modeling
of the light variation provides yet another strong argument that
OGLE-LMC-T2CEP-290 and OGLE-LMC-T2CEP-291 are first-overtone
pulsators.

The increased mass has its consequence, however. Evolution of single
stars does not lead to such massive horizontal branch stars. Although
evolution in a binary system would provide a solution, higher mass
implies higher luminosity, which is inconsistent with
observations. More extensive modeling is planned to resolve this
puzzle.

\acknowledgments

This work has been supported by the National Science Centre, Poland,
grant MAESTRO no. 2016/22/A/ST9/00009. The OGLE project has received
funding from the Polish National Science Centre grant MAESTRO
no. 2014/14/A/ST9/00121. IS and RS are supported by the Polish
National Science Centre grant OPUS no. DEC-2015/17/B/ST9/03421.

\end{document}